\documentclass[%
 reprint,
nofootinbib,
nobibnotes,
bibnotes,graphicx,graphics,
 amsmath,amssymb,
 aps,stmaryrd,
]{revtex4-1}
\usepackage{natbib} 
\usepackage{xtab,afterpage}
\usepackage{amsfonts}
\usepackage{soul}
\usepackage{tikz}

\usepackage{xtab,afterpage}
\usepackage{nicematrix}
\usepackage{graphicx}

\newcommand{\comment}[1]{}
\usepackage{tikz}
  \newlength\squareheight
\usepackage{wasysym} 
\usepackage{graphicx}
\usepackage{dcolumn}
\usepackage{bm}

\begin{document}

\preprint{Draft}

\title{Mixing patterns in graphs with higher-order structure}

\author{Peter Mann}
\email{pm78@st-andrews.ac.uk}
\author{Lei Fang}
\author{Simon Dobson}
\affiliation{School of Computer Science, University of St Andrews, St Andrews, Fife KY16 9SX, United Kingdom }

\date{\today}

\begin{abstract}
In this paper we examine the percolation properties of higher-order networks that have non-trivial clustering and subgraph-based assortative mixing (the tendency of vertices to connect to other vertices based on subgraph joint degree). Our analytical method is based on generating functions. We also propose a Monte Carlo graph generation algorithm to draw random networks from the ensemble of graphs with fixed statistics. We use our model to understand the effect that network microstructure has, through the arrangement of clustering, on the global properties. Finally, we use an edge disjoint clique cover to represent empirical networks using our formulation, finding the resultant model offers a significant improvement over edge-based theory. 
\end{abstract}

\pacs{Valid PACS appear here}
\maketitle


\section{Introduction}
\label{sec:introduction}

Traditional network theory is limited to pairwise interactions. Higher-order, non-dyadic interactions lead to fresh insight into collective dynamics that cannot be described by edge-based theory \cite{PhysRevLett.124.218301,iacopini_petri_barrat_latora_2019,giusti_ghrist_bassett_2016,XXXXXXXJJJJJJ}. Theoretical models that include higher-order interactions become increasingly complicated compared to edge-based models and generalisations of the definitions of traditional concepts, such as assortativity or clustering are often required.

The ability to generate a random graph from an ensemble of random graphs that have statistically equivalent properties allows us to model, and subsequently to understand, the topological features that drive the behaviour of dynamical processes that occur over networks. Topological features can be extrinsic (finite-size effects), whilst others are intrinsic, such as the clustering coefficient, $\mathcal C$ \cite{PhysRevE.68.026121,PhysRevLett.103.058701}, or the degree assortativity \cite{PhysRevE.67.026126,PhysRevLett.89.208701,balogh_palla_kryven_2020}. Some properties are locally defined, and as such are subject to local anisotropy; whilst others characterise global attributes of the network that are averaged over all vertices. Finally, ensemble properties manifest for a collection of graphs that are statistically equivalent in some manner. Often, the intrinsic properties of networks are not independent of one another. For instance, two networks with equivalent clustering coefficients could exhibit different responses to a given dynamical process due to the organisation of their triangles. Clustering is known to increase the critical point of the bond percolation process and reduce the size of the expected giant connected component (GCC) \cite{PhysRevE.80.020901,gleeson_2009,PhysRevE.83.056107,PhysRevE.81.066114,PhysRevE.104.024303,PhysRevE.103.012313,PhysRevE.101.062310,PhysRevE.103.012309,PhysRevE.104.024304}. However, vertices involved in many triangles that tend to connect to other vertices with a high triangle count may induce assortative correlations among the overall degrees. In turn, the critical point of positively assorted networks is reduced compared to the neutral model \cite{PhysRevE.67.026126} and other properties can also be affected \cite{PhysRevE.67.027101}. The observed behaviour of a percolation process depends on the interplay between the clustering and the assortativity \cite{PhysRevLett.103.058701}. Therefore, it is important to model both the role of clustering as well as the role of degree assortativity.

\begin{figure}[h]
    \centering
    \includegraphics[width=0.32\textwidth]{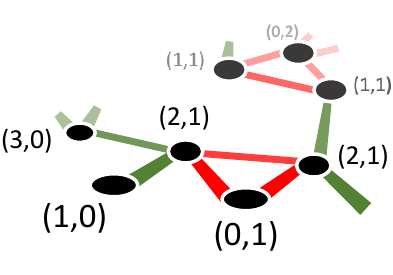}
    \caption{A snapshot of a 2- and 3-clique clustered graph with vertices labelled with their joint degree tuples $(s,t)$.}    
    \label{fig:joint_degrees_image}
\end{figure}

One modelling framework is the configuration model (CM) that enables the construction of a particular locally tree-like random graph from a distribution of degrees \cite{newman_strogatz_watts_2001}. CM networks are regarded to be infinitely large, such that fluctuations from ensemble averages do not effect the statistics of the model. This model has received significant attention in the literature, and has been extended to the so-called generalised configuration model (GCM) to incorporate non-trivial clustering and higher-order structure among vertices \cite{PhysRevE.82.066118}. In the GCM, a set $\vec \tau$ of substrate motifs is defined (such as cliques or chordless cycles of different sizes) and the number of edge-disjoint motifs of each topology that a given vertex belongs to is described by a tuple, known as its joint degree that takes values in $\mathbb Z_{\geq 0}^n$ where $n$ is the number of distinct motif topologies in the model. For example, for a model containing only 2- and 3-cliques, a vertex that belongs to 3 ordinary edges, 2 triangles has joint degree $(s,t)=(3,2)$, see Fig \ref{fig:joint_degrees_image}. Vertex joint degrees are drawn from a multivariate distribution of joint degrees, $p_{s,t,h,\dots}$, that defines the probability that a vertex belongs to $s$ ordinary edges, $t$ triangles, $h$ 4-cycles etc. When $\vec \tau$ is chosen to contain only ordinary edges (2-cliques), the GCM collapses to the standard CM. 
The CM is known to create networks with no assortativity among the degrees of neighbouring vertices due to the stochastic nature of the construction algorithm \cite{PhysRevE.82.066118}; vertex degrees are independent random variables. Assortativity can be inserted, in a controlled manner, into the CM via a degree-preserving Markov Chain Monte Carlo (MCMC) rewiring algorithm where the correlations converge to a target value \cite{PhysRevLett.89.208701,PhysRevE.67.026126}. 

Hasegawa \textit{et al} \cite{PhysRevE.101.062310} examined the role of clustering and assortativity in the percolating component of tree-triangle networks. Recently, Mann \textit{et al} investigated the inter-subgraph correlation properties that arise naturally in the percolating component of GCM networks that are composed of arbitrary sized cliques that have neutral mixing patterns in the substrate graph \cite{PhysRevE.105.044314}. 

In this paper, we extend that work to examine the inter-subgraph properties of GCM networks that have arbitrary clique clustering and subgraph correlations in the substrate network. To do this, we develop an analytical method based on generating functions and introduce an MCMC rewiring algorithm to simulate such networks. These steps allow the synthesis of random graphs drawn from an ensemble of graphs with a given clustering coefficient $\mathcal C$ and correlation structure among the joint degree tuples. This model can be used to probe the role that both clustering and assortativity have on the properties of the bond percolation process and will likely lead to the detailed study of higher-order structure for other kinds of dynamical processes on networks beyond the percolation model. Additionally, our model allows the clustering and correlation structure from an empirical network to be gathered through means of an edge-disjoint clique cover \cite{burgio_arenas_gomez_matamalas_2021, PhysRevE.105.044314}. Through our MCMC algorithm, synthetic networks, possibly containing additional vertices to the empirical network, can then be created. In this way, our model allows ensemble representations with fixed statistics of empirical networks to be created and studied. We remark, that the information in our model can readily be compressed from inter-subgraph correlations to correlations between overall degrees; however, there is a corresponding information loss following this process. 

\section{Theoretical}
\label{sec:theory}

Correlations can arise via three distinct processes in GCM networks. Firstly, correlation between the joint degrees of neighbouring vertices that belong to the same clique type. For instance, three vertices in a triangle must all have $t\neq 0$. In this manner, vertices with membership of a given 3-clique are somewhat assorted together; they will never connect to vertices that don't belong to triangles. However, this correlation does not extend beyond the scope of the motif to which the edge belongs and so, we refer to it as the \textit{trivial correlation}. In other words, the excess joint degrees (the remaining joint degree tuples excluding the motif to which the edge belongs) of two vertices at the ends of a randomly selected edge are not correlated in the default GCM construction. And, excluding the trivial assortativity due to belonging to the same motif, there are no specific correlations between vertices at play. 

We can also discuss assortativity between the overall degrees of vertices in the GCM. The overall degree of a vertex that belongs to $s$ 2-cliques, $t$ 3-cliques, $w$ 4-cliques (and so on) is given by $k = s+2t+3c+\dots$. This assortativity often arises in an uncontrolled manner. For example, a GCM network composed of 2-cliques and 10-cliques may be absent of any correlation structure among the distribution in the numbers of cliques a vertex belongs to; however, by virtue of the increase in overall degree, vertices that belong to 10-cliques will likely be of higher-degree than vertices that belong to just 2-cliques, and as a result, will exhibit positive correlations among the overall degrees if this is the case \cite{PhysRevE.80.020901,PhysRevE.101.062310}. 

Finally, we can inject correlation into GCM networks deliberately, in a detailed and controlled fashion, by specifying how likely it is that a vertex with a given joint degree tuple $(s_0,t_0,\dots,c_0)$ has a neighbour with a certain joint degree tuple $(s',t',\dots,c')$. In general, this probability is different depending on the topology of the edge (to which clique it belongs to). This kind of correlation is the most general of the three types that arise in the GCM; since the other correlation types are subsets of this; and so, it is the focus of this paper.

\begin{figure}[h]
    \centering
    \includegraphics[width=0.495\textwidth]{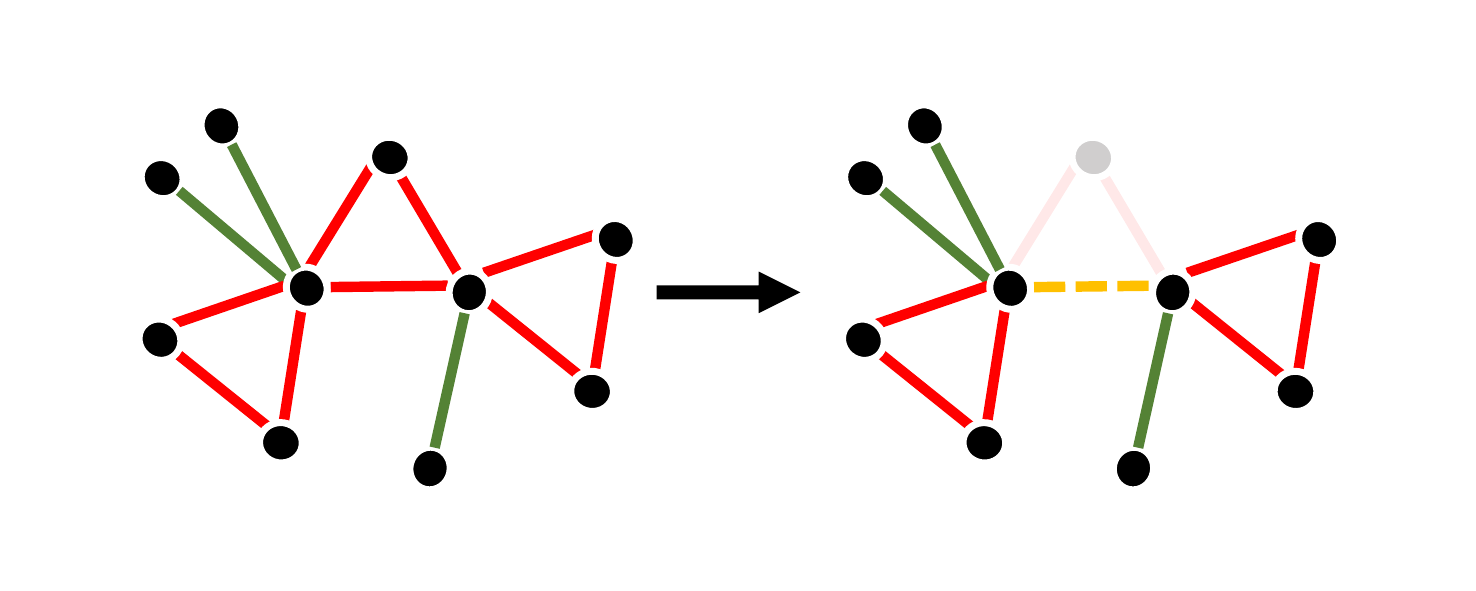}
    \caption{The joint excess joint degree distribution of topology $\tau$ is created by selecting all $\tau$-edges and measuring the joint excess degrees of the terminal vertices, excluding the motif the selected edge belongs to. In the example, the selected edge (yellow dashed) has vertices of excess joint degrees $(2,1)$ and $(1,1)$.}    
    \label{fig:joint_excess_joint_degree}
\end{figure}

We restrict our attention to GCM networks that are composed of 2- and 3-cliques, symbolised by $\vec\tau=\{\bot,\Delta\}$, respectively.
Let $e_{\tau, s_0t_0,s't'}$ be the probability that a randomly chosen $\tau$-edge leads to vertices with excess (remaining) joint degrees of $(s_0,t_0)$ and $(s',t')$ where $s$ and $t$ are the numbers of 2- and 3-cliques, respectively, that each vertex belongs to, see Fig \ref{fig:joint_excess_joint_degree}. The joint excess joint degree distribution is the data input to the model; other familiar quantities such as excess joint degree distribution, $q_{\tau,s,t}$ and the joint degree distribution, $p_{s,t}$ can be derived from these information-rich function. 

There is an $e_{\tau, s_0t_0,s't'}$ function for each $\tau \in \vec \tau$; since, we can follow $\text{card}(\vec \tau)$ distinct edge-topologies. We assume that
\begin{equation}
   \sum_{s_0}\sum_{t_0}\sum_{s'}\sum_{t'}e_{\tau, s_0t_0,s't'}=1,
\end{equation}
with all indices running from $0,\dots,\infty$ unless otherwise stated. We can recover the excess joint degree distribution, which is the probability that an edge of topology $\tau$ is chosen at random and leads to a vertex of remaining degree $(s_0,t_0)$, as
\begin{equation}
    q_{\tau,s_0t_0} = \sum_{s'}\sum_{t'}e_{\tau, s_0t_0,s't'}\label{eq:excess}.
\end{equation}
We can invert $q_{\tau,s,t}$ to obtain $p_{st}$ for each $\tau\in\{\bot,\Delta\}$. For the tree-triangle model this is given by 
\begin{subequations}
\begin{align}
    p^{\bot}_{s,t} =& \frac{\bigg[q_{\bot,s-1,t}\big/s\bigg]}{\bigg[\sum\limits_{s't'}q_{\bot,s'-1,t}\big/s'\bigg]}\label{eq:pbot},\\
    p^{\Delta}_{s,t} =& \frac{\bigg[q_{\Delta,s,t-1}\big/t\bigg]}{\bigg[\sum\limits_{s't'}q_{\Delta,s',t-1}\big/t'\bigg]},\label{eq:ptriangle}
\end{align}
\end{subequations}
which can be verified by inserting $q_{\bot,s,t} = {(s+1)p_{s+1,t}}/{\langle s\rangle}$ into Eq \ref{eq:pbot} or $q_{\Delta,s,t} = {(t+1)p_{s,t+1}}/{\langle t\rangle}$ into Eq \ref{eq:ptriangle} where $\langle s\rangle$ is the average number of 2-cliques a vertex belongs to 
\begin{equation}
    \langle s\rangle =\sum_s\sum_t s p_{s,t},
\end{equation}
and similarly for $\langle t\rangle$. We remark that, in general, the inverted distributions, $p_{s,t}^\tau$, are not numerically equivalent to one another, nor do they necessarily contain all $(s,t)$ tuples. This is because, the excess joint degree distributions are not equivalent to one another for different topologies. To see this, $q_{\bot,s,t}$, (the joint excess distribution obtained from randomly selected 2-clique edges) can yield no information on the marginal distribution of how vertices with zero 2-cliques, but non-zero 3-clique counts interact; and vice versa for $q_{\Delta,s,t}$; since, we cannot follow a 2-clique edge to a vertex with no 2-cliques. However, the actual joint degree distribution, $p_{s,t}$, can be recovered by multiplying each partially observed joint degree distribution by the fraction of the network (by vertex count) that was used to create the observation. For instance, $p_{s,t} = n_\tau p^\tau_{s,t}$ where $n_\tau$ are the fraction of vertices that belong to at least one $\tau$ motif. For a coherent model, we have $ n_\tau p^\tau_{s,t}= n_\nu p^\nu_{s,t}$ for a given $(s,t)$ joint degree and for $\tau,\nu\in \vec\tau$. Each $n_\tau$ can readily be found as one minus the fraction of vertices that do not belong to any $\tau$ cliques; for instance $n_\bot = 1 - \sum_tp_{0,t}$. When all vertices belong to both 2- and 3-cliques, $n_\tau=n_\nu$, and the inverted joint distributions coincide.
Finally, it should be noted that we can obtain the joint excess degree distribution between overall degrees $e_{j,k}$ from $e_{\tau,s_0t_0,s't'}$ by application of a Kronecker delta
\begin{align}
    e_{j,k} =& \sum_{s_0,t_0}\sum_{s't'}w_\bot e_{\bot,s_0t_0,s't'}\delta_{j,s_0+2t_0}\delta_{k,s'+2t'}\nonumber\\
    &+\sum_{s_0,t_0}\sum_{s't'}w_\Delta e_{\Delta,s_0t_0,s't'}\delta_{j,s_0+2t_0+1}\delta_{k,s'+2t'+1}.\label{eq:collapse}
    \end{align}
The $+1$ terms in the final two delta functions account for the contribution of the other edge within the 3-clique towards the overall degree. The weighting factors $w_\tau$ account for the required renormalisation from the independently normalised distributions to a single distribution. These are given by 
\begin{equation}
    w_\bot = \frac{\sum\limits_{st}sp_{st}}{\sum\limits_{st}sp_{st} + 2\sum\limits_{st}tp_{st}},
\end{equation}
and 
\begin{equation}
    w_\Delta = \frac{2\sum\limits_{st}tp_{st}}{\sum\limits_{st}sp_{st} + 2\sum\limits_{st}tp_{st}}.
\end{equation}
This expression compresses the information available from the model and collapses to the formulation derived in \cite{PhysRevLett.89.208701}. Additionally, the overall degree distribution $p_k$ can readily be found from a joint degree distribution \cite{PhysRevLett.103.058701}
\begin{equation}
    p_k=\sum_{st}p_{st}\delta_{k,s+2t}.
\end{equation}

\subsection{$G_0$ generating function}
\label{sec:treetrianglegeneratingfunction}

We can use $e_{\tau,s_0t_0,s't'}$ to find ensemble properties by deriving a generating function formulation for the model. Suppose that a vertex of joint degree $(s_0,t_0)$ is chosen from the network. Now suppose that we traverse the $s_0$ 2-clique edges and record the excess joint degrees $(s',t')$ of those neighbouring vertices. Let $\{r_\bot\}=\{r_{\bot,s_1,t_1},\dots,r_{\bot,s_n,t_n}\}$ be the configuration of those neighbour joint degrees we might record; such that there are $r_{\bot,s',t'}$ neighbours of excess joint degree $(s',t')$. The probability of a given configuration of excess joint degrees surrounding the focal vertex following $s_0$ 2-clique edges is
\begin{widetext}
\begin{equation}
    P(\{r_{\bot}\}\mid s_0) = s_0!\prod_{s,t}\frac{1}{r_{\bot,st}!}\left(\frac{e_{\bot,s_0-1,t_0,s,t}}{\sum\limits_{s,t}e_{\bot,s_0-1,t_0,s,t}}\right)^{r_{\bot,st}}.
\end{equation}
Similarly, let us traverse each of the $2t_0$ 3-clique edges from the focal vertex to the neighbours along the corners of each triangle and record the excess joint degree of those vertices. We record the number of each unique excess joint degree tuple as $r_{\Delta,s't'}$ to find the configuration $\{r_\Delta\}$ surrounding the focal vertex along all of its 3-clique edges. The probability of a given configuration is the multinomial
\begin{equation}
    P(\{r_{\Delta}\}\mid t_0) = 2t_0!\prod_{s,t}\frac{1}{r_{\Delta,st}!}\left(\frac{e_{\Delta,s_0,t_0-1,s,t}}{\sum\limits_{s,t}e_{\Delta,s_0,t_0-1,s,t}}\right)^{r_{\Delta,st}}.
\end{equation}
The probability that a vertex with joint degree $(s_0,t_0)$ has a neighbour configuration $\{r_\bot\}$ and $\{r_\Delta\}$ is simply the product of these terms
\begin{equation}
    P(\{r_{\bot}\}, \{r_{\Delta}\}\mid s_0, t_0) = P(\{r_{\bot}\}\mid s_0)P(\{r_{\Delta}\}\mid t_0)\delta\left(s_0,\sum\limits_{s,t}r_{\bot,st}\right)\delta\left(2t_0,\sum\limits_{s,t}r_{\Delta,st}\right),
\end{equation}
where $\delta(i,j)$ is the Kronecker delta that ensures only configurations that sum to $s_0$ and $2t_0$ 2- and 3-clique edges, respectively are retained. We can then generate this expression by summing over all of the possible configurations along each edge topology
\begin{equation}
    \mathcal P_{s_0, t_0}(z_\bot,z_\Delta) =  \sum_{\{r_{\bot}\}}\sum_{\{r_{\Delta}\}}P\left(\{r_{\bot}\}, \{r_{\Delta}\}\mid s_0, t_0\right)\left(\prod_{s,t}z_{\bot,st}^{r_{\bot,st}}\right)\left(\prod_{s,t}z_{\Delta,st}^{r_{\Delta,st}}\right)\label{eq:Ps0t0zz},
\end{equation}
where 
\begin{equation}
    \sum_{\{r_{\tau}\}} := \sum_{r_{\tau,s_1t_1}}\sum_{r_{\tau,s_1t_2}}\cdots \sum_{r_{\tau,s_{n}t_{m}}},
\end{equation}
and where $s_n$ $(t_m)$ is the largest 2-clique (3-clique) degree. The arguments of $P_{s_0,t_0}$ are vectors containing an element for each $(s',t')$ pair in $e_{\tau,s_0,t_0,s't'}$ for each clique topology $\{\bot,\Delta\}$ such that $z_\tau = z_{\bot,s_1t_1},\dots, z_{\bot,s_nt_m}$. We then apply the multinomial theorem
\begin{equation}
    (z_1+z_2+\dots+z_m)^n=\sum_{k_{1}}\sum_{k_{2}}\cdots\sum_{k_{m}}\delta\left(\sum\limits_{t}^m k_t,n\right)\frac{n!}{k_1!k_2!\cdots k_m!}\prod_{t=1}^mz_t^{k_t}\label{eq:multinomial},
\end{equation}
to Eq \ref{eq:Ps0t0zz} and average over the probability of choosing a vertex of joint degree $(s_0,t_0)$ to arrive at a generating function for a randomly selected vertex
\begin{subequations}
\begin{align}
    \mathcal G_0(z_\bot,z_\Delta^2) =& \sum_{s_0}\sum_{t_0}p_{s_0t_0} \left(\frac{\sum\limits_{s,t}e_{\bot,s_0-1,t_0,s,t}z_{\bot,st}}{\sum\limits_{s,t}e_{\bot,s_0-1,t_0,s,t}}\right)^{s_0}\left(\frac{\sum\limits_{s,t}e_{\Delta,s_0,t_0-1,s,t}z_{\Delta,st}}{\sum\limits_{s,t}e_{\Delta,s_0,t_0-1,s,t}}\right)^{2t_0},\\
    =&G_0\left(\frac{\sum\limits_{s,t}e_{\bot,s_0-1,t_0,s,t}z_{\bot,st}}{\sum\limits_{s,t}e_{\bot,s_0-1,t_0,s,t}},\left[\frac{\sum\limits_{s,t}e_{\Delta,s_0,t_0-1,s,t}z_{\Delta,st}}{\sum\limits_{s,t}e_{\Delta,s_0,t_0-1,s,t}}\right]^2
    \right),
\end{align}
\end{subequations}
with $G_0(x,y)=\sum_s\sum_tp_{st}x^sy^{t}$. This is the fundamental generating function of the model to obtain the properties of randomly selected vertices. We can extract the properties of individual joint degrees from this distribution as well as vertices with overall degree $k$, irrespective of their joint degree by multiplying by a delta  $G_0(z) = G_0(z,z^2)= \sum_kp_kz^k=\sum_k\sum_{st}p_{st}z^{s+2t}\delta_{k,s+2t}$. In this case our expressions reduce to those found by Newman \cite{PhysRevE.67.026126}.

We now calculate the size $S$ of the GCC by introducing a generating function for each joint excess joint degree distribution. For this we need to generate the probability of following an edge of a given topology to a vertex and that vertex failing to belong to the GCC. For this purpose, let
\begin{align}
      \mathcal F_{\tau,s_0t_0} (z_{\bot}, z_{\Delta})&= \frac{\sum\limits_{s'}\sum\limits_{t'}e_{\tau,s_0t_0,s't'}z_{\bot,s't'}^{s'}z^{t'}_{\Delta,s't'}}{\sum\limits_{s'}\sum\limits_{t'}e_{\tau,s_0t_0,s't'}}.
\end{align}
We define $u_{\bot,s_0,t_0}$ to be the probability that a 2-clique connected
to a vertex of remaining degree $(s_0,t_0)$ at one end leads to another vertex that does not belong to the giant component. Similarly, we define $u_{\Delta,s_0,t_0}$ to be the probability that a randomly selected 3-clique edge that has one vertex of joint excess degree $(s_0,t_0)$ fails to connect to the GCC via the other end of the edge. We can write self-consistency expressions for these probabilities by summing over all possible neighbouring vertex excess joint degrees multiplied by the probability that their edges fail to connect that vertex as 
\begin{subequations}
\begin{align}
    u_{\bot,s_0,t_0} =& \mathcal F_{\bot,s_0t_0}(g_{\bot},g^2_{\Delta}),  \\
    u_{\Delta,s_0,t_0} =& \mathcal F_{\Delta,s_0t_0}(g_{\bot},g^2_{\Delta}) ,
\end{align}\label{eq:jointF}
\end{subequations}
where $g_\bot$ and $g_\Delta$ are vectors with elements $g_{\bot,s,t}=1-\phi+u_{\bot,s,t}\phi$ and $g_{\Delta,s,t}^2=(1-\phi+u_{\Delta,s,t}\phi)^2-2\phi^2(1-\phi)u_{\Delta,s,t}(1-u_{\Delta,s,t})$, which are the probabilities that a 2-clique edge and both edges in a 3-clique fail to connect to the GCC \cite{PhysRevE.80.020901}, respectively for bond occupancy probability $\phi$. 
The GCC is then found from
\begin{equation}
    S=1-G_0\left(\frac{\sum\limits_{s,t}e_{\bot,s_0-1,t_0,s,t}g^s_{\bot,st}g_{\Delta,s+1,t-1}^{2(t-1)}}{\sum\limits_{s,t}e_{\bot,s_0-1,t_0,s,t}},\left[\frac{\sum\limits_{s,t}e_{\Delta,s_0,t_0-1,s,t}g_{\bot,s-1,t+1}^{s-1}g^{2t}_{\Delta,st}}{\sum\limits_{s,t}e_{\Delta,s_0,t_0-1,s,t}}\right]^2
    \right).\label{eq:main_23_GCC}
\end{equation}
Note, some the indices contain $s\pm 1$ and $t\mp 1$; since, given an excess 2-clique joint degree, $(s,t)$, we obtain the excess 3-clique joint degree as $(s+1,t-1)$; similarly, given an excess 3-clique joint degree of $(s,t)$, we obtain the 2-clique excess joint degree as $(s-1,t+1)$. We extend this model to GCM networks comprised of larger clique sizes in Appendix \ref{sec:theory:largercliques}.

\subsection{Percolation threshold}
\label{sec:treetrianglethreshold}

Like other percolation models on random graphs, the model exhibits a 2nd order critical point in $\phi$ with order parameter $S$. This indicates that the order parameter is continuous at the transition, but its derivative is not. The value of $\phi$ at the critical point, $\phi_{\text{crit}}$, can be found by perturbing around the point $u_\tau=1$ $\forall \tau\in \vec \tau$, which corresponds to the probability that all edges, regardless of topology, fail to lead to the GCC. As $\phi\gtrsim \phi_{\text{crit}}$ each $u_\tau$ drops below 1 and we can expand both expressions in Eqs \ref{eq:jointF} using a Taylor series \cite{PhysRevLett.103.058701, 10.1371/journal.pone.0071321} about $z_{\tau,st} = 1-u_{\tau,st}$ to obtain
\begin{align}
    z_{\tau,s_0t_0} \approx\ &
    1-\mathcal F_{\tau,s_0t_0}(\bm 1,\bm 1) - \sum_s\sum_t\frac{\partial \mathcal F_{\tau,s_0t_0}}{\partial g_{\bot,st}}\frac{\partial g_{\bot,st}}{\partial z_{\bot,st}}\bigg|_{z_{\bot,st}=0}-2\sum_s\sum_t\frac{\partial \mathcal F_{\tau,s_0t_0}}{\partial g_{\Delta,st}}\frac{\partial g_{\Delta,st}}{\partial z_{\Delta,st}}\bigg|_{z_{\Delta,st}=0}\nonumber,
    \\
    \approx\ &
    \phi\sum_s\sum_t \mathcal F^{(\bot,st)}_{\tau,s_0t_0}(\bm 1,\bm 1)z_{\bot,st}+2\phi(1+\phi-\phi^2)\sum_s\sum_t\mathcal F_{\tau,s_0t_0}^{({{\Delta,st}})}(\bm 1,\bm 1)z_{\Delta,st}+(\mathcal O^2),
\end{align}
\end{widetext}
where $\mathcal F_{\tau,st}^{{\nu,st}}$ is the partial derivative of $\mathcal F_{\tau,st}$ with respect to $g_{\nu,st}$ and where we have treated $g_{\tau,st}(z_{\tau,st})$ with $g_{\tau,st}(0)=1$. We have assumed that $\mathcal F_{\tau,st}$ is well-behaved at $u=1$, with finite derivatives, which does not hold for power law degree distributions. The system can be expressed as a matrix equation $\bm z=\bm {\mathcal F}
\bm \phi\cdot \bm z$ where $\bm {\mathcal F}$ is the Jacobian of $\mathcal F_{\tau,st}$ $\forall \tau\in \{\bot,\Delta\}$, $\bm z$ is the vector $(z_{\bot,s_1t_1},\dots,z_{\Delta,s_nt_m})$ and 
\begin{equation}
    \bm\phi = \sum_{i=1}\mathbf{c}_i^T\mathbf{g}\mathbf{c}_i\mathbf{c}_i^T,
\end{equation}
where $\mathbf{g}=(g'_{\bot,s_1t_1},\dots,g'_{\Delta,s_nt_m})$ is the vector of derivatives of $g_{\tau,st}(z_{\tau,st})$ at $z_{\tau,st}=0$, $\mathbf{c}_i$ is the i-th basis vector of $\mathbb{R}^{n_\bot m_\bot+n_\Delta m_\Delta}$ and $T$ denotes the transpose. For a given $\phi$ the system exhibits non-trivial solution for $\bm z$ when $\det(\bm {\mathcal F} \bm \phi-\bm I)=0$ where $\bm I$ is the identity matrix. The scalar value $\phi_{\text{crit}}$ (that uniquely determines matrix $\bm \phi$) that satisfies this expression can then be found from the characteristic equation.

\subsection{Joint correlation functions}
\label{sec:Correlationfunctions}

Traditionally, the information of the overall degree correlation properties of a network is summarised by an assortativity coefficient, $r^{\text{Pearson}}_k$. This can be obtained from the family of $e_{\tau,st,s't'}$ matrices by first finding $e_{j,k}$ via Eq \ref{eq:collapse} before 
evaluating 
\begin{equation}
    r^{\text{Pearson}}_k = \frac{1}{\sigma^2_{q_{k}}}\sum_{jk}jk(e_{jk} - q_jq_k),
\end{equation}
where 
\begin{equation}
    \sigma^2_{q_k} = \sum\limits_k k^2q_{k} - \left(\sum\limits_k kq_{k}\right)^2,
\end{equation}
is the variance of the excess degree distribution of overall degrees $q_{k}$, which in turn, is given by the summation of the rows $ q_{k} = \sum_{j}e_{jk}$ \cite{PhysRevE.67.026126}. In general we expect a well-behaved assortativity coefficient to take its values in $r\in[-1,1]$, with positive (negative) coefficient indicating assortative (dissasortative) mixing and $r=0$ indicating random mixing. Finally we note that the expected statistical error in an assortativity coefficient can be calculated by the jackknife method \cite{efron_1979} or the bootstrap method \cite{10.1214/ss/1177013815}. 

In this section, we will examine possible extensions of the assortativity coefficient to encapsulate the mixing properties of vertices based on joint degree tuples. A natural definition is to simply extend the Pearson correlation coefficient along each edge topology. For instance, if a network comprises of 2- and 3-cliques, then we obtain two assortativtiy coefficients; one for each $\tau\in \vec \tau=\{\bot,\Delta\}$ given by
\begin{equation}
    r^{\text{Pearson}}_\tau = \frac{1}{\sigma_{q_\tau}^2}\sum_{s,t,s',t'}sts't'(e_{\tau,st,s't'} - q_{\tau,st}q_{\tau,s't'}),\label{eq:r_tau_pearson}
\end{equation}
where $\sigma_{q_\tau}$ is the standard deviation of the excess joint degree distribution $q_{\tau,st}$ in topology $\tau$, normalising the function by its maximum value
\begin{equation}
    \sigma^2_{q_{\tau}} = \sum\limits_{s,t} ({st})^2q_{\tau,st} - \left(\sum\limits_{st} stq_{\tau,st}\right)^2.
\end{equation}
The square of the standard deviation (the variance) is strictly positive unless all vertices have the same joint degree, in which case the coefficient is undefined. 

Let us now introduce an additional assortativity coefficient that classifies joint degree tuples independently and considers the inter-class mixing properties. In this case the assortativity coefficient along each edge topology is 
\begin{equation}
    r_\tau^{\text{class}} = \frac{\sum\limits_{st}e_{\tau,st,st} - \sum\limits_{st} \left(\sum\limits_{s't'}e_{\tau,st,s't'}\right)\left(\sum\limits_{s't'}e_{\tau,s't',st}\right)}{1-\sum\limits_{st} \left(\sum\limits_{s't'}e_{\tau,st,s't'}\right)\left(\sum\limits_{s't'}e_{\tau,s't',st}\right)}.\label{eq:r_tau_trace}
\end{equation}
This assortativity measure examines the extent to which each class of joint degree tuple mixes with itself in the network. More compactly, and with generalisation to models with additional clique sizes, we write \cite{PhysRevE.67.026126,PhysRevE.99.042306}
\begin{equation}
    r_\tau^{\text{class}} = \frac{\text{Tr}[e_\tau]-||{e_\tau^2}||}{1-||{e_\tau^2}||},\label{eq:assortativity_coefficient}
\end{equation}
where $||x||$ is the sum of all elements of matrix $x$ and $\text{Tr}[x]$ is the trace.
This information is collected into a tuple of coefficients $r^{\text{class}}=(r^{\text{class}}_\bot,r^{\text{class}}_\Delta,\dots,r^{\text{class}}_\gamma)$ that describe the mixing patterns along each motif topology. Defining the coefficient in this way is equivalent to mapping each distinct joint degree tuple to a colour and then comparing the mixing patterns of vertex colours for the different edge types \cite{balogh_palla_kryven_2020}. The drawback of this coefficient is the loss of a metric between pairs of non-identical joint degrees. For example, considering joint degree tuples that take the form $(s,t)$, this definition would not find the joint degree $(1,2)$ any more similar to $(1,3)$ than to $(10,20)$, despite the obvious difference in connectivity. This classification-based coefficient is therefore stricter than the degree-based one; since, it can exhibit no correlation, even when overall degrees are correlated or when either of the marginal dimensions are correlated. 

There are other ways to define a assortativity coefficients \cite{PhysRevE.87.022801}; especially when there are vector attributes placed on vertices \cite{,pelechrinis_wei_2016,rabbany_eswaran_dubrawski_faloutsos_2017}. And, whilst there is not a universal approach in the network science literature, most methods embed the vector quantity in a metric space and compare distances along each component. We leave further examination of assortativity coefficients for future work.

\subsection{Choosing $e_{\tau,st,s't'}$ matrices}
\label{sec:matrices}

In order to use the formulation, we must find suitable $e_{\tau,st,s't'}$ matrices for each edge type $\tau\in \{\bot,\Delta\}$ and the choice will determine the mixing properties of the model. By far the simplest approach is to simply extract these relationships from empirical networks, and we discuss this in section \ref{sec:discussion}. In this section, we overview manual processes instead.

A matrix $e_{\tau, st,s't'}$ is a valid if it satisfies the following conditions
\begin{subequations}
\begin{align}
    e_{\tau, st,s't'} &\geq 0\\
    e_{\tau, st,s't'} &= e_{\tau, s't',st}\\
    \sum_{s't'}e_{\tau, st,s't'}&=q_{\tau,st},\\
    \sum_{st}q_{\tau,st} &=1,
\end{align}
\label{eq:constraints}
\end{subequations}
with $q_{\tau,st}\in [0,1]$. In addition, we require the further constraint that the inversion of matrices $e_{\tau, st,s't'}$ and $e_{\nu, st,s't'}$ for $p_{st}$ through Eqs \ref{eq:pbot} and \ref{eq:ptriangle} coincide, thus constraining the excess joint degree distributions $q_\tau$ and $q_\nu$.

For neutral mixing, in the absence of any inter-subgraph assortativity, the joint excess joint degree distribution is given by the product of the excess joint degrees
\begin{equation}
    e^{(\text{n})}_{\tau,st,s't'}=q_{\tau,st}q_{\tau,s't'}.\label{eq:neutralmixing}
\end{equation}

For a perfectly assorted network, in which edges only connect vertices with the same excess joint degree we have
\begin{equation}
    e^{(\text{a})}_{\tau,st,s't'} = q_{\tau,st}\delta_{s,s'}\delta_{t,t'}.
\end{equation}
It is convenient to parameterise the extent of the assortativity with an infinitesimal value $\epsilon$. We can readily construct these matrices from the neutral case by re-distributing the row summations (excess joint degree distribution) such that the diagonal is given by
\begin{equation}
    e^{(\text{a})}_{\tau,st,st} = \sum_{s't'}e_{\tau,st,s't'} - \epsilon\sum_{s't'}q_{\tau,st}q_{\tau,s't'}(1-\delta_{s,s'}\delta_{t,t'}),\label{eq:assortative1}
\end{equation}
whilst off-diagonal entries, $s\neq s'$ and $t\neq t'$ are given by
\begin{equation}
    e^{(\text{a})}_{\tau,st,s't'} =\epsilon\sum_{s't'}q_{\tau,st}q_{\tau,s't'}(1-\delta_{s,s'}\delta_{t,t'}).\label{eq:assortative2}
\end{equation}

For a perfectly dissasortative network, in which a vertex with a given joint degree will never connect to a neighbour with the same joint degree we have  
a zero diagonal element
\begin{equation}
    e^{(\text{d})}_{\tau,st,st} = 0,\qquad \forall (s,t)
\end{equation}
The off-diagonal elements are given by a linear combination of the row sums whose coefficients are $\pm 1$. The row sum $q_i$ and column sum $q_j$ for element $e_{ij}$ always appears with a positive sign; whilst, the coefficients of the other row and column sums are such that they sum to zero for any given row or column. For instance, consider the following 3x3 example
\begin{equation}
e^{(\text{d})}_{\tau}
=
\renewcommand*{\arraystretch}{1.5}
\begin{pmatrix}
0 & \frac{q_1+q_2-q_3}{2} & \frac{q_1-q_2+q_3}{2} \\
\frac{q_1+q_2-q_3}{2} & 0 & \frac{-q_1+q_2+q_3}{2} \\
\frac{q_1-q_2+q_3}{2} & \frac{-q_1+q_2+q_3}{2}  & 0 \\
\end{pmatrix},\label{eq:3x3}
\end{equation}
where $q_i$ corresponds to the sum of elements on row $i$. The elements are divided by 2 since there are 2 non-zero values to distribute the row sum over per row. We present more details in Appendix \ref{appendix:B}.

We can relax the strict zero diagonal property of these matrices by introducing infinitesimal values $\epsilon_i$ along the diagonal such as
\begin{equation}
e^{({\text{r}})}_{\tau}
= \epsilon + e^{(\text{d})}_{\tau}
\end{equation}
where $ \epsilon $ is the matrix 
\begin{equation}
 \epsilon 
=
\renewcommand*{\arraystretch}{1.5}
\begin{pmatrix}
\epsilon_1 & \frac{-\epsilon_1-\epsilon_2+\epsilon_3}{2} & \frac{-\epsilon_1+\epsilon_2-\epsilon_3}{2} \\
\frac{-\epsilon_1-\epsilon_2+\epsilon_3}{2} & \epsilon_2 & \frac{\epsilon_1-\epsilon_2-\epsilon_3}{2} \\
\frac{-\epsilon_1+\epsilon_2-\epsilon_3}{2} & \frac{\epsilon_1-\epsilon_2-\epsilon_3}{2}  & \epsilon_3 \\
\end{pmatrix}.\label{eq:3x32}
\end{equation}

\section{Simulation}

\begin{figure*}[t]
    \centering
    \includegraphics[width=1.0\textwidth]{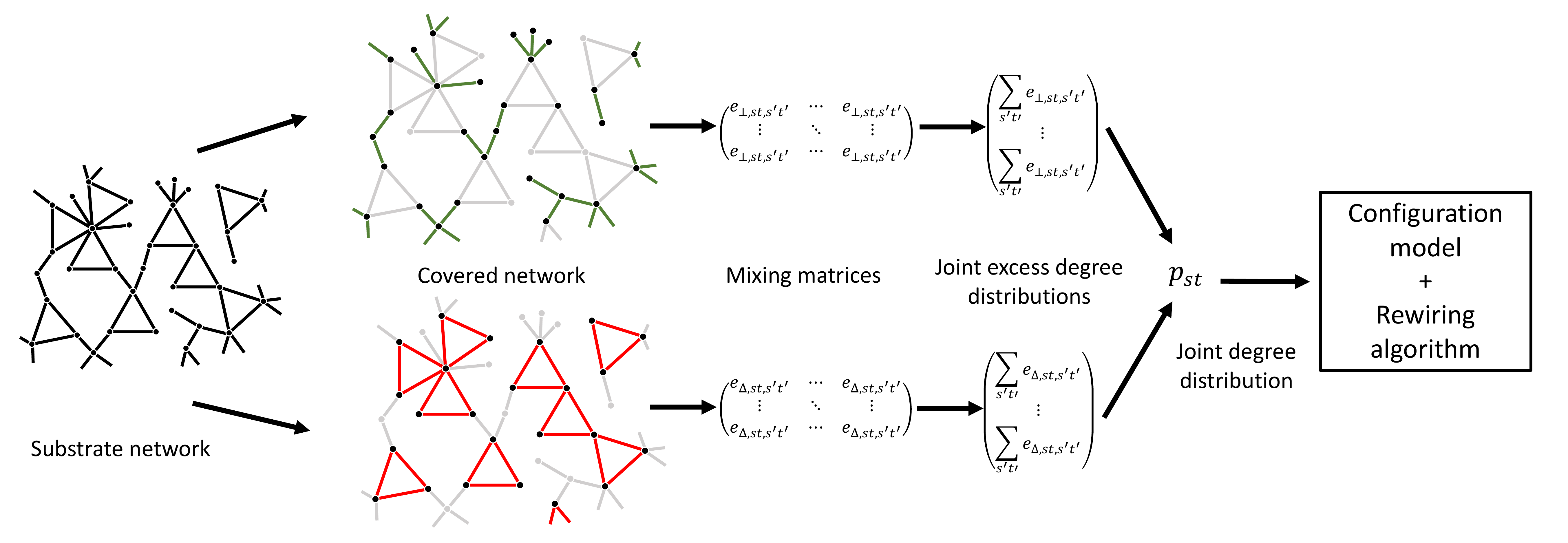}
    \caption{An overview, from left to right, of the workflow to create a random graph of arbitrary size with statistically equivalent clustering and inter-clique mixing properties from a substrate network. A substrate network is covered by an edge-disjoint clique cover before the mixing matrices are extracted along each edge type. The excess joint degree distributions can then be calculated and inverted to find the joint degree distribution $p_{st}$. Random networks can then be created using the configuration model and the resulting synthetic network can be rewired by our MCMC stochastic algorithm. }    
    \label{fig:workflow}
\end{figure*}

With a theoretical framework in place to study the percolation properties of clustered random graphs that exhibit a given mixing pattern, we would now like to discuss how to simulate such networks. The data input to the simulation framework is a coherent set of $e_{\tau}$ matrices for a set of permissible subgraph topologies $\vec\tau$. We have two options: i) extract $e_{\tau}$ from an empirical network; or ii) write $e_{\tau}$ analytically. We have discussed how to generate matrices analytically in section \ref{sec:matrices} and so, we now overview a workflow to examine empirical networks. The workflow is summarised in Fig \ref{fig:workflow} and defined as follows. For a given empirical graph, an edge-disjoint (non-overlapping) clique cover can be used to cover the network by a predetermined basis set of clique subgraphs. Once the cover has been applied to the network, the joint excess joint degree distribution, $e_{\tau,{s,t,\dots},{s',t',\dots}}$ can be gathered; from which, the excess joint degree distributions, $q_{\tau,{s,t,\dots}}$ and the joint degree distribution $p_{s,t,\dots}$ can be found. Random graphs can be created according to the GCM prescription and these graphs can be post-processed using a stochastic MCMC rewiring algorithm, which we introduce in Section \ref{sec:MCMC}, to yield the correct mixing patterns.

Given the centrality of covering empirical networks with an edge-disjoint set of cliques to the application of this work, it is appropriate to comment on existing methods in the literature. To the author's knowledge, there are two distinct edge-disjoint clique covers: i) the EECC \cite{burgio_arenas_gomez_matamalas_2021} developed by Burgio \textit{et al} and ii) the MPCC \cite{PhysRevE.105.044314} developed by Mann \textit{et al}. Both of these covers are stochastic heuristics that cover a network in cliques up to some tunable maximum size such that a given edge in the substrate network belongs to a single clique. The EECC optimises on the number of cliques that are used to cover the graph, aiming for as few as possible, such a cover is a \textit{maximal cover}. Conversely, the MPCC gives precedence to larger cliques, aiming to retain as much local structure as possible for high-degree vertices and neighbourhoods. Details on these algorithms can be found in \cite{burgio_arenas_gomez_matamalas_2021,PhysRevE.105.044314}.

Once we have covered the empirical network in motifs, we can then extract the joint excess joint degree distributions, $e_{\tau,{s,t,\dots},{s',t',\dots}}$ from it; subsequently, the excess joint degree distributions, $q_{\tau,{s,t,\dots}}$ and the joint degree distribution $p_{s,t,\dots}$ can be obtained. Random graphs $G(V,E)$ can now be created according to the GCM algorithm by drawing vertex joint degrees from  $p_{s,t,\dots}$ to create a sequence of joint degrees of length $|V|$ \cite{PhysRevE.82.066118,PhysRevE.103.012313,PhysRevE.103.012309}. We note that the number of vertices $|V|$ is independent of the size of the empirical network, and so, this method can be used to enlarge network datasets for subsequent analysis whilst holding the clustering coefficient and the motif distribution fixed. At this point, the inter-motif correlations are not fixed. We then create a list $l_\tau$ for each distinct clique size $\tau\in\vec\tau$ and enter the identity of each vertex once into those lists for each distinct motif that it belongs to of size $\tau\in \vec \tau$. The lists of vertex identities are then randomised. We now construct the network. For each clique we draw without replacement the appropriate number of vertices from each list; equivalently, we partition the list into segments with length equal to the number of vertices for a given size clique. We repeat this for all motifs until the lists are exhausted; all edges have been placed in the network and the algorithm is terminated.

In general, it is highly unlikely that the same joint degree sequence will be drawn from $p_{s,t,\dots}$ or that the identity of the vertices that belong to a motif will be identical upon repeat of this algorithm. This means that we will obtain a different graph $G'(V,E)$ that belongs to the same ensemble of random graphs $\mathcal G$ that share equivalent statistical properties. We remark that the accidental selection of the same vertex from a clique list, or the accidental combining of one or more edges between motifs (such that they are not edge-disjoint) is possible; however, both events are increasingly uncommon as $|V|$ increases for a fixed joint degree distribution and so will not effect the statistics of the model.

\begin{figure}[h]
    \centering
    \includegraphics[width=0.375\textwidth]{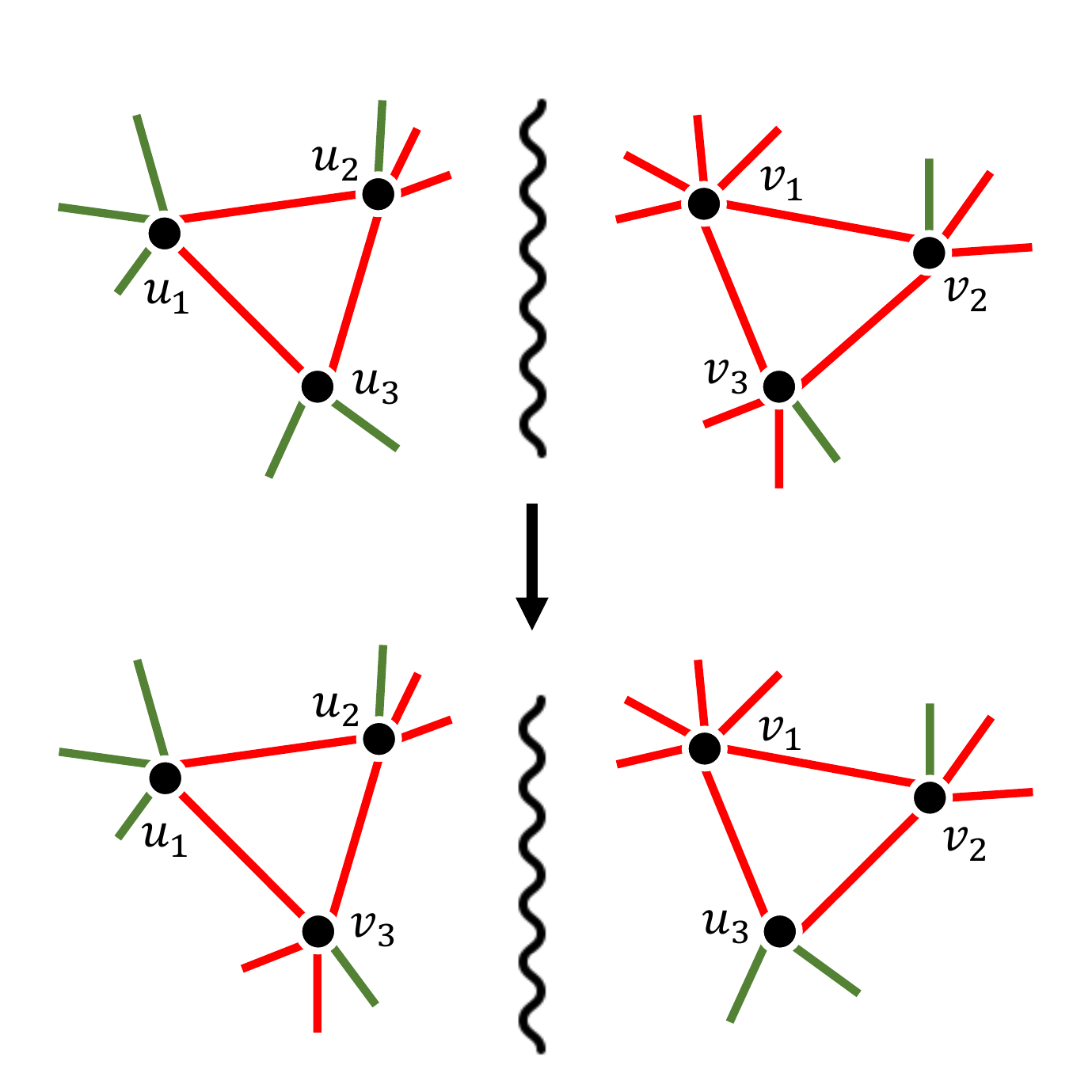}
    \caption{The rewiring process for two triangle corners in the network. Edges $(u_1,u_3),(u_2,u_3),(v_1,v_3)$, and $(v_2,v_3)$ are broken and new edges $(u_1,v_3),(u_2,v_3),(v_1,u_3)$ and $(v_2,u_3)$ are created.}    
    \label{fig:mcmc}
\end{figure}

\label{sec:MCMC}
 The graphs created via the GCM are absent of correlations between motifs; although they may display overall degree correlations due to the sizes of each clique. To inject a correlation structure between the motifs, we introduce an MCMC rewiring algorithm whose target distribution yields the correct $e_{\tau}$ joint excess joint degree distribution for all $\tau\in \vec \tau$ that appear in the model. At each stage of the algorithm a clique size is selected from $\vec \tau$. Next, two vertices that belong to cliques of that size are selected at random from different motifs in the network. The edges that connect the vertices to the clique are broken. The two vertices are exchanged and new edges are created to rewire the incident vertices to their new locations. For example, in Fig \ref{fig:mcmc}, a network comprising 2- and 3-cliques is rewired according to our MCMC algorithm. In this case, vertices $u_3$ and $v_3$ are swapped to yield an alternative configuration that preserves the joint degree sequence. To achieve this, edges $(u_1,u_3),(u_2,u_3),(v_1,v_3)$, and $(v_2,v_3)$ are broken and edges $(u_1,v_3),(u_2,v_3),(v_1,u_3)$ and $(v_2,u_3)$ are created. The proposed swap is then accepted with probability $P=\text{min}(1,\pi)$ where
\begin{equation}
    \pi = \frac{e_{\Delta,k_{u_1},k_{v_3}}e_{\Delta,k_{u_2},k_{v_3}}e_{\Delta,k_{v_1},k_{u_3}}e_{\Delta,k_{v_2},k_{u_3}}}{e_{\Delta,k_{u_1},k_{u_3}}e_{\Delta,k_{u_2},k_{u_3}}e_{\Delta,k_{v_1},k_{v_3}}e_{\Delta,k_{v_2},k_{v_3}}}
\end{equation}
where $k_{i}=(s_{i},t_{i})$ in $e_{\tau,k_i,k_j}$ is the excess joint degree of vertex $i$ following topology $\tau$. This Metropolis-Hastings condition is the quotient of the product of the proposed joint excess joint degree distributions following the swap and their non-permuted counterparts.
For larger clique sizes, when vertex $u_1$ is swapped with $v_1$, the Metropolis-Hastings acceptance condition is written 
\begin{equation}
    \pi=\prod_i\frac{e_{\tau,k_{u_1},k_{v_i}}e_{\tau,k_{v_1},k_{v_i}}}{e_{\tau,k_{u_1},k_{u_i}}e_{\tau,k_{v_1},k_{v_i}}}
\end{equation}
where index $i$ runs over the number of edges being broken in each clique (i.e. its size minus 1).

It is clear to see that this dynamics conserves the joint degree sequence and is ergodic on the set of graphs having that joint degree sequence. By this, we mean that any graph with the joint degree sequence can be obtained in a finite number of Monte Carlo updates. The algorithm also satisfies detailed balance for any graph configuration $\lambda$ that occurs with probability  
\begin{equation}
    p_\lambda = \prod_{\tau}\prod_{i=1}^{M_{\tau}}e_{\tau_i}
\end{equation}
where the product is over all distinct clique sizes $\tau$ in the network and all edges that belong to cliques of that size $i=1,\dots,M_{\tau}$ in the graph. Detailed balance can be shown by considering the transition probability between any two states $\lambda\rightarrow\lambda'$ by evaluating $ p_\lambda P(\lambda\rightarrow\lambda') = p_{\lambda'} P(\lambda'\rightarrow\lambda)$.
With ergodicity and detailed balance satisfied, the MCMC algorithm yields a graph that has the required correlation structure as its fixed point.

\section{Discussion}
\label{sec:discussion}

\subsection{$k$-regular graphs}
\label{sec:discussion:treetriangle}

Consider a random graph ensemble that consists of networks that have both 2- and 3-clique subgraphs. Let the joint degree distribution $p_{st}$ be given by $(5,1)=1/3,\ (3,2)=1/3$ and $(1,3)=1/3$ such that the global clustering coefficient and overall vertex degree are fixed for all networks ($k$-regular).
In this example, by following 2-clique edges, we can reach vertices with joint excess degrees $(0,3),\ (4,1)$ and $(2,2)$; i.e. these are the excess joint degrees in $q_{\bot,
st}$. Similarly, by following triangle edges to a corner, we can reach vertices with joint excess degrees $(1,2),\ (0,5)$ and $(3,1)$; which constitute the excess joint degrees of $q_{\Delta,st}$. 

The mixing matrices $ e_\tau^{\text{neutral}} $ for the uncorrelated model are given by 
\begin{equation}
\newcommand{\rh}[1]{\rotatebox[origin=c]{60}{$\scriptstyle#1$}}
\newcommand{\uh}[1]{\vcenter{\hbox{$\scriptstyle#1$}}}
\NiceMatrixOptions{%
    code-for-first-row = \scriptstyle,
    code-for-first-col = \scriptstyle,
                    }
\renewcommand\arraystretch{1.5}
e_{\bot}^{(\text{n})} = 
\begin{pNiceArray}[first-row,first-col]{ccc}
        &       \rh{(0,3)}   &       \rh{(4,1)}   &       \rh{(2,2)}   \\
(0,3)   &    \frac{1}{81} &  \frac{5}{81} &   \frac{3}{81}\\
(4,1)   &    \frac{5}{81} &  \frac{25}{81} & \frac{15}{81}\\
(2,2)   &    \frac{3}{81}  &  \frac{15}{81} & \frac{9}{81} \\
\end{pNiceArray}
\end{equation}
and
\begin{equation}
\newcommand{\rh}[1]{\rotatebox[origin=c]{60}{$\scriptstyle#1$}}
\newcommand{\uh}[1]{\vcenter{\hbox{$\scriptstyle#1$}}}
\NiceMatrixOptions{%
    code-for-first-row = \scriptstyle,
    code-for-first-col = \scriptstyle,
                    }
\renewcommand\arraystretch{1.5}
 e_\Delta^{(\text{n})} = 
\begin{pNiceArray}[first-row,first-col]{ccc}
        &       \rh{(3,1)}   &       \rh{(1,2)}   &       \rh{(5,0)}   \\
(3,1)   &    \frac{16}{144} & \frac{24}{144} & \frac{8}{144} \\
(1,2)   &    \frac{24}{144} & \frac{36}{144} & \frac{12}{144} \\
(5,0)   &    \frac{8}{144} & \frac{12}{144} & \frac{4}{144} \\
\end{pNiceArray}
\end{equation}
For example, element $e^{(\text{n})}_{\bot,(0,3),(0,3)}=1/9\times 1/9$ since $1/9$ of all tree edges lead to a vertex with joint degree $(1,3)$. Similarly, element $e^{(\text{n})}_{\Delta,(3,1),(3,1)}=4/12\times 4/12$ since four of the 12 3-clique edges lead to vertices with joint excess degree $(3,1)$. 

\begin{figure}[h]
    \centering
    \includegraphics[width=0.45\textwidth]{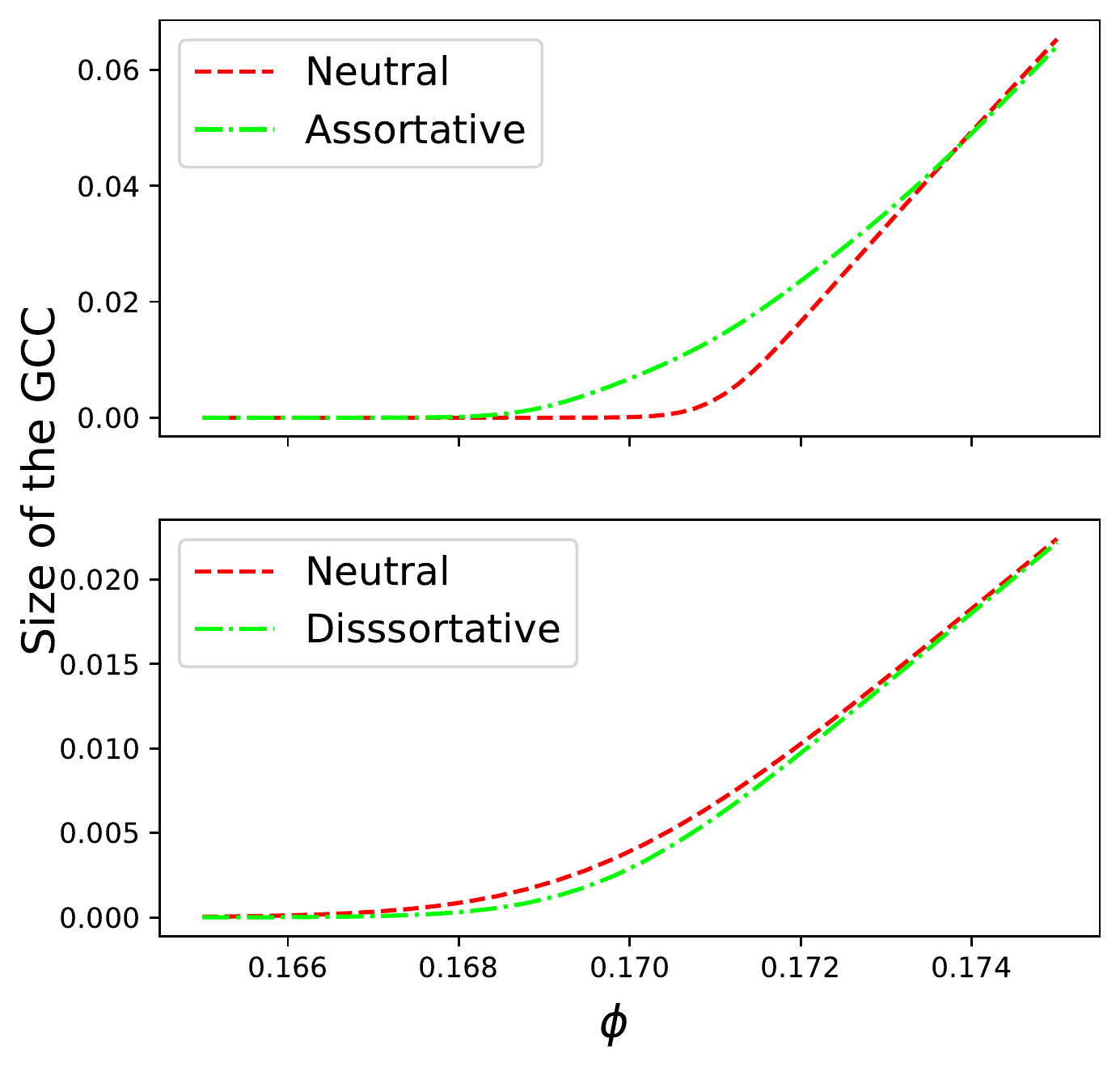}
    \caption{The expected size of the GCC for neutral and correlated random networks in $k$-regular graphs with 2- and 3-clique clustering among vertices with overall degree $s+2t=7$ and fixed clustering coefficients. Curves are the theoretical results of Eq \ref{eq:main_23_GCC}.}
    \label{fig:neutral_dis_toy}
\end{figure}

We can make assorted matrices $e_\tau^{(\text{a})}$ by pushing the probability mass into the diagonal entries and setting the off-diagonal terms to a infinitesimal value, according to Eqs \ref{eq:assortative1} and \ref{eq:assortative2}.
\begin{widetext}
\begin{equation}
\newcommand{\rh}[1]{\rotatebox[origin=c]{60}{$\scriptstyle#1$}}
\newcommand{\uh}[1]{\vcenter{\hbox{$\scriptstyle#1$}}}
\NiceMatrixOptions{%
    code-for-first-row = \scriptstyle,
    code-for-first-col = \scriptstyle,
                    }
\renewcommand\arraystretch{1.5}
e_\bot^{(\text{a})}  = 
\begin{pNiceArray}[first-row,first-col]{ccc}
        &       \rh{(0,3)}   &       \rh{(4,1)}   &       \rh{(2,2)}   \\
(0,3)   &    \frac{9}{81}-\epsilon\frac{5}{81}-\epsilon\frac{3}{81} & \epsilon\frac{5}{81}  & \epsilon\frac{3}{81}\\
(4,1)   &    \epsilon\frac{5}{81} & \frac{45}{81}-\epsilon\frac{5}{81}-\epsilon\frac{15}{81} & \epsilon\frac{15}{81} \\
(2,2)   &    \epsilon\frac{3}{81} & \epsilon\frac{15}{81} & \frac{27}{81}-\epsilon\frac{3}{81}-\epsilon\frac{15}{81} \\
\end{pNiceArray}
\end{equation}

\begin{equation}
\newcommand{\rh}[1]{\rotatebox[origin=c]{60}{$\scriptstyle#1$}}
\newcommand{\uh}[1]{\vcenter{\hbox{$\scriptstyle#1$}}}
\NiceMatrixOptions{%
    code-for-first-row = \scriptstyle,
    code-for-first-col = \scriptstyle,
                    }
\renewcommand\arraystretch{1.5}
e_\Delta^{(\text{a})}  = 
\begin{pNiceArray}[first-row,first-col]{ccc}
        &       \rh{(3,1)}   &       \rh{(1,2)}   &       \rh{(5,0)}   \\
(3,1)   &    \frac{48}{144}-\epsilon\frac{24}{144}-\epsilon\frac{8}{144} & \epsilon\frac{24}{144}  & \epsilon\frac{8}{144}\\
(1,2)   &    \epsilon\frac{24}{144} & \frac{72}{144}-\epsilon\frac{24}{144}-\epsilon\frac{12}{144} & \epsilon\frac{12}{144} \\
(5,0)   &    \epsilon\frac{8}{144} & \epsilon\frac{12}{144} & \frac{24}{144}-\epsilon\frac{8}{144}-\epsilon\frac{12}{144} \\
\end{pNiceArray}
\end{equation}
\end{widetext}
where $0\lesssim\epsilon $. This procedure establishes an equivalence class between the excess joint degree distributions of the assorted and neutral correlation matrices; and consequently, between the inverted joint degree distribution. In other words, the joint degree distribution and the excess joint degree distributions, the clustering coefficients and the overall degree of the vertices of the networks that can be created will all be equivalent; however, their inter-subgraph mixing patterns exhibit different structure. 
The expected size of the GCC for the neutral and assorted mixing matrices is shown in the top plot of Fig \ref{fig:neutral_dis_toy}. Despite being $k$-regular, assortativity between the joint degrees changes the percolation properties of the network, reducing the critical point.

Next, we examine the effect of disassortative mixing in a $k$-regular network with joint degree distribution $p_{st}$ given by $(5,1)\approx0.282$, $(3,2)\approx0.374$ and $(1,3)\approx0.344$. The neutral mixing matrices are given by Eq \ref{eq:neutralmixing} whilst the dissasorted matrices are given by Eq \ref{eq:3x3}. In Fig \ref{fig:neutral_dis_toy}, we plot the results of Eq \ref{eq:main_23_GCC} for the two mixing patterns, observing that inter-subgraph mixing leads to an (albeit small) increase in the critical point of the model, in line with traditional theory of dissasortative mixing between overall degrees \cite{PhysRevLett.89.208701}. However, due to the vertices having the same overall degrees, other theoretical models could not exhibit this behaviour. Importantly, our model provides evidence that assortativity plays an important role in the observed percolation dynamics, even when the clustering coefficient is fixed (in addition to the overall degree); settling a longstanding debate on the subject \cite{PhysRevE.80.020901,PhysRevE.98.062314,PhysRevE.81.066114}. 

\subsection{Empirical graphs}
For our next analysis, we examine the correlation properties of a series of empirical networks. The datasets are the network of hyperlinks between weblogs on US politics \cite{10.1145/1134271.1134277}; a network of coauthorships between scientists posting preprints on the Condensed Matter E-Print Archive between 1995 and 1999 \cite{doi:10.1073/pnas.98.2.404}; the Astrophysics collaboration network \cite{konect:2017:ca-AstroPh,konect:leskovec107}; Twitter \cite{NIPS2012_7a614fd0}; friendship data of Facebook users \cite{konect:2017:facebook-wosn-links,viswanath09}; the social network of Douban \cite{konect:2017:douban,konect:socialcomputing}; user–user friendship relations from Brightkite \cite{konect:2017:loc-brightkite_edges,konect:cho2011}; the collaboration graph of authors of scientific papers from the arXiv's Astrophysics (astro-ph) section \cite{konect:2017:ca-AstroPh,konect:leskovec107};  the network of connections between autonomous systems of the Internet \cite{konect:2017:topology,konect:zhang05}; the network of autonomous systems of the Internet connected with each other from the CAIDA project \cite{konect:2017:as-caida20071105,konect:2017:as-caida20071105}; coappearance network of characters in the novel Les Miserables \cite{knuth_knuth_knuth_2009}; the C. Elegans network \cite{watts_strogatz_1998}.

\begin{figure*}[t]
    \centering
    \includegraphics[width=1.0\textwidth]{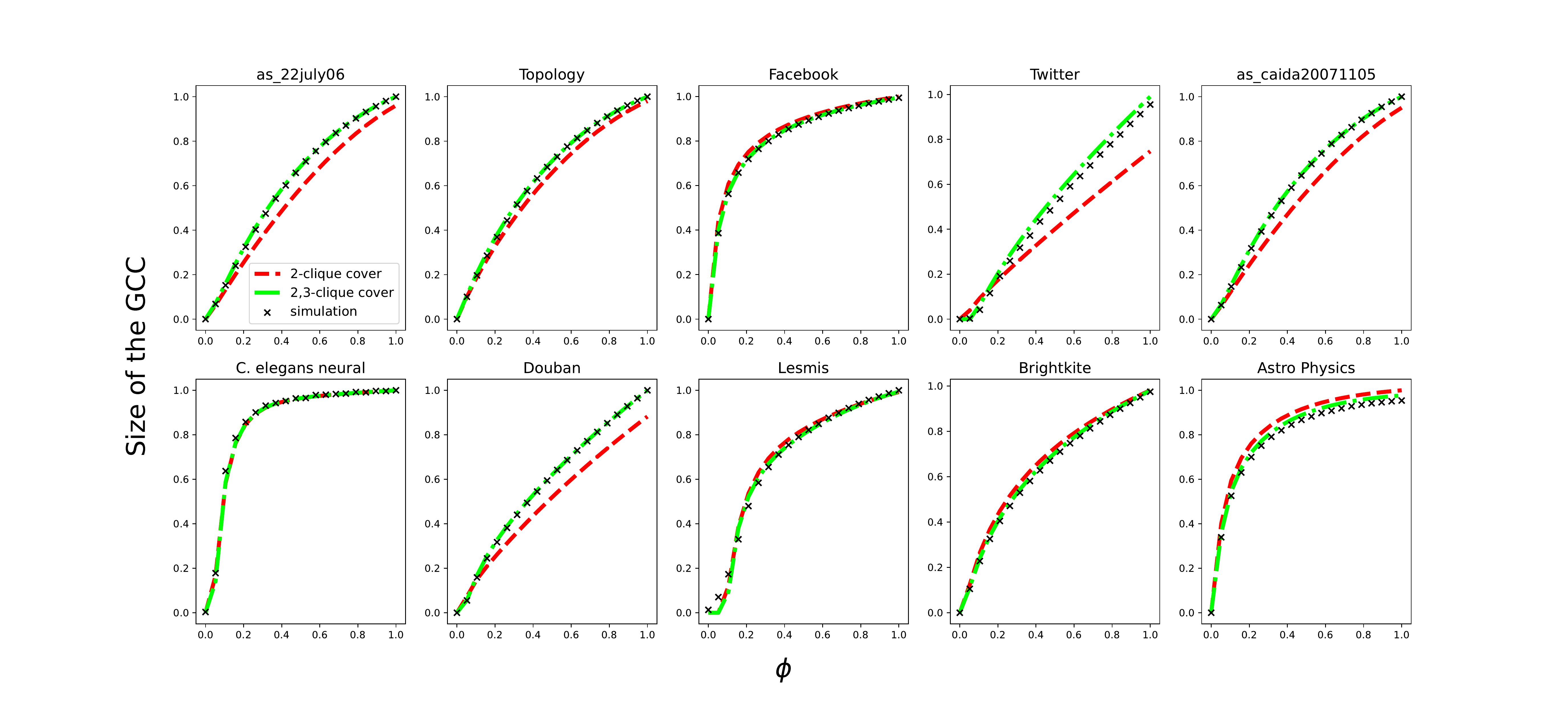}
    \caption{The size of the GCC for a series of empirical network datasets. The scatter points are the average of Monte Carlo bond percolation on the original network. The (red) dashed lines are the expected theoretical curves when using a 2-clique approximation \cite{PhysRevLett.89.208701} ; whilst the (green) dashed and dotted curves are the results for the 2- and 3-clique theory of Eq \ref{eq:main_23_GCC}. In each case, including higher-order interactions from the empirical network yields a better theoretical representation.
    }    
    \label{fig:size_gcc}
\end{figure*}

In Fig \ref{fig:size_gcc}, we plot the size of the GCC for a series of empirical networks under different clique covers. Scatter points are the result of Monte Carlo simulations of bond percolation on the original graph; whilst the (red) dashed lines are the theoretical results of traditional edge-based theory (Eq 9 of \cite{PhysRevLett.89.208701}) for 2-clique covered networks; finally, the (green) dashed and dotted lines are the results of 2- and 3-clique covered networks in combination with Eq \ref{eq:main_23_GCC}. Across the series, it is immediately evident that the higher-order mixing properties are crucial in capturing the percolation response of the empirical network. In each case, our 2-3 clique model is in strong agreement with the dynamics of the empirical graph. It is interesting that in some case, such as C. Elegans and the LesMis network, that the 2-clique theory is in close agreement with the simulation; whilst, for social networks, such as the Twitter and Douban networks, the edge-based theory offers poor agreement with simulation. 

\begin{figure*}[t]
    \centering
    \includegraphics[width=1.0\textwidth]{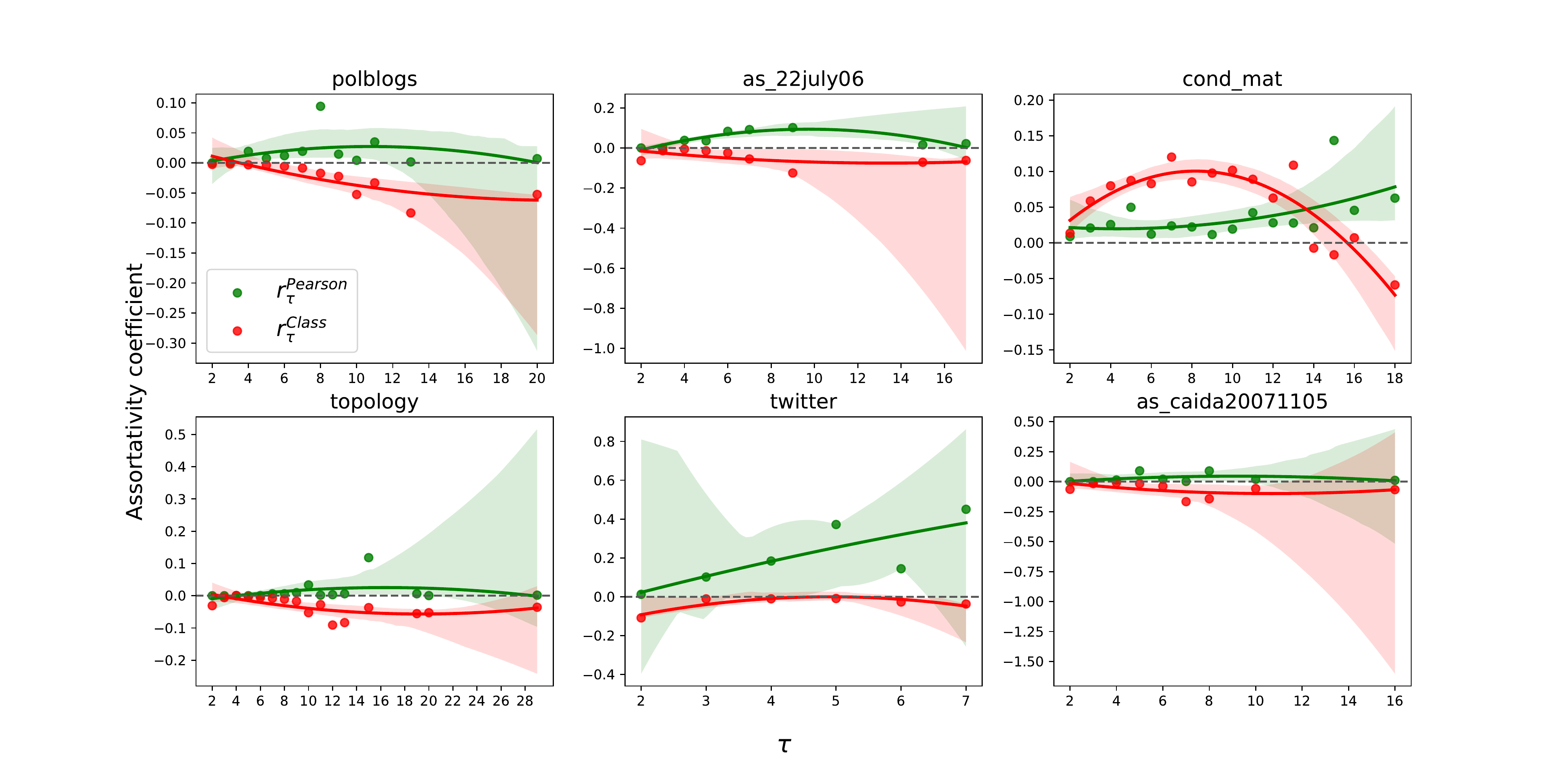}
    \caption{Scatter plots of the assortativity coefficients $r_\tau^{\text{Pearson}}$ and $r_\tau^{\text{Class}}$ given by Eqs \ref{eq:r_tau_pearson} and  \ref{eq:r_tau_trace}, respectively of 6 empirical networks with a linear regression model fit. The shaded regions are the bootstrap confidence interval for the regression estimate whilst the markers are the data points. To collect the data, the networks were covered using an MPCC before the $e_\tau$ mixing matrices were extracted and used to calculate the assortativity coefficient along each edge topology.
    }    
    \label{fig:assortativtiy_all_cliques}
\end{figure*}
In Fig \ref{fig:assortativtiy_all_cliques} we analyse the inter-subgraph mixing properties for a range of empirical networks. For each dataset we first find a full MPCC cover of the network before extracting the mixing matrices $e_{\tau}$. These matrices are then used to calculate each $r_\tau^{\text{Pearson}}$ and $r_\tau^{\text{Class}}$ through Eqs \ref{eq:r_tau_pearson} and \ref{eq:r_tau_trace}, respectively. Scatter points are the values of the empirically obtained correlation coefficients for each clique size, whilst lines are fitted regression plots with shaded confidence intervals. We observe a range of mixing patterns in the networks, in particular we notice that vertices that belong to cliques of size $2<c<7$ tend to exhibit assortative mixing. For example, the increase in the Pearson coefficient for vertices in the Twitter network. However, larger clique sizes tend to be dissasortatively mixed, for instance, cliques of size larger than $10$ in the condensed matter coauthorship network. This is almost certainly due to the smaller number of larger cliques in these finite sized networks.

\section{Conclusion}

In this paper we have introduced a theoretical framework, including a generating function formulation and rewiring algorithm, that accounts for higher-order mixing patterns in networks with clustering. We have applied our framework to random graph ensembles, conceptual models and to empirical graphs by use of edge-disjoint clique covers from the literature. Members (graphs) within the ensembles have the following fixed statistics: joint degree distribution, clustering coefficient and inter-motif correlation structure. We found that our model offers excellent agreement with Monte Carlo simulation and yields a significant improvement beyond the results of ordinary edge-based theory. We found that non-trivial correlation structure of empirical networks as a function of cliques size, indicating the presence of structure at different scales. 

We have provided an analytical description of how to write the input data to this model, $e_{\tau}$ matrices as well as how to mine them from empirical networks. To simulate such networks we have introduced an MCMC rewiring algorithm that, in conjunction with the well-known GCM algorithm, yields synthetic networks with the statistically equivalent mixing properties as the empirical models. 

This method can be used to draw samples of synthetic network datasets with fixed statistical properties that are representative of the original data in terms of clustering and mixing patterns. Our formulation is focused around clique subgraphs; however, we anticipate the generalisation of this work to hypergraphs in future work, and we believe this would offer an increasingly realistic model of empirical networks and greater understanding of their properties. We anticipate the application of our model to a variety of dynamical processes and networked datasets, which we leave for future work.

For subgraphs larger than 2-cliques, our method inherently captures beyond nearest-neighbour connections within the inter-subgraph correlation structure. However, it falls short of being a sufficient model of long range inter-subgraph correlations; a feature that would allow higher-order equivalence classes among tensor quantities that capture the $n$-hop correlation structures to be written.  By increasing the rigidity of the local environment around a vertex, the degrees of freedom of the GCM model and consequently the number of possible graphs in the random graph ensemble is reduced. Properties such as connectedness, core-periphery sub-structure, beyond nearest-neighbour inter-subgraph correlations and modularity are crucial mesoscale aspects of network topology whose control would reduce the number of configurations within an ensemble of random graphs, and therefore yield models that are increasingly representative of empirical data.    

The role of correlations of this nature has implications for the spread of epidemic diseases, and their non-trivial thresholds, and should be studied in this context \cite{PhysRevE.103.062308,PhysRevE.103.042307}. We leave this for future work.

\section{ACKNOWLEDGMENTS}

This work was partially supported by the UK Engineering and Physical Sciences Research Council under grant number EP/N007565/1 (Science of Sensor Systems Software).


\bibliography{ref}

\appendix

\section{larger cliques}
\label{sec:theory:largercliques}

We now generalise the model developed in  section \ref{sec:treetrianglegeneratingfunction} to account for GCM networks that contain larger clique subgraphs beyond 2- and 3-cliques. Suppose that a vertex of joint degree $(s_0,t_0,\dots,c_0)$ is chosen from a random graph that is comprised of edge-disjoint cliques up to some maximum size $C$. Now suppose that we traverse each of the edges of a given clique size $\tau$ and record the excess joint degrees $(s',t',\dots,c')$ of the neighbouring vertices and then repeat this for all $\tau\in \vec \tau$. Let $\{r_\tau\}=\{r_{\tau,s_1,t_1,\dots,c_1},\dots,r_{\tau,s_n,t_n,\dots,c_n}\}$ be the configuration of those neighbour joint degrees we might record; such that there are $r_{\tau,s',t',\dots,c'}$ neighbours of excess joint degree $(s',t',\dots,c')$ and so on. The probability of a given configuration of excess joint degrees surrounding the focal vertex is 
\begin{widetext}
\begin{align}
     \mathcal P_{s_0,\dots,c_0} =&\ p_{s_0\dots c_0}\sum_{\{r_\bot\}}\cdots\sum_{\{r_C\}}\prod_\tau (\tau-1)k_0!\prod_{s,\dots,c}\frac{1}{r_{\tau,s\dots c}!}\nonumber\\
     &\times\left(\frac{e_{\tau,s_0,\dots,k_0-1,\dots,c_0,s,\dots,k,\dots,c} }{\sum\limits_{s,\dots,c}e_{\tau,s_0,\dots,k_0-1,\dots,c_0,s,\dots,k,\dots,c}}z_{\tau,s\dots c}\right)^{r_{\tau,s\dots c}}\delta\left((\tau-1)k_0,\sum\limits_{s,\dots,c}r_{\tau,s\dots c}\right)
\end{align}
and after application of Eq \ref{eq:multinomial} we have 
\begin{align}
    \mathcal P_{s_0,\dots,c_0}=p_{s_0\dots c_0} \prod_{\tau}\left(\frac{\sum\limits_{s,\dots,c}e_{\tau,s_0,\dots,k_0-1,\dots,c_0,s,\dots,k,\dots,c}z_{\tau,s,\dots,c}}{\sum\limits_{s,\dots,c}e_{\tau,s_0,\dots,k_0-1,\dots, c_0,s,\dots,c}}\right)^{(\tau-1)k_0}
\end{align}
which yields the average properties of the first order neighbours of a vertex with joint degree $(s_0,\dots,c_0)$ that has been randomly selected from the network. This is an information-rich distribution function from which we can extract the average properties $\mathcal P_k$ of vertices with overall degree $k$ by filtering out those joint degrees that sum to $k$ as
\begin{equation}
    \mathcal P_{k} = \sum_s\dots \sum_c \mathcal P_{s\dots c} \delta\left(k,\sum_\tau (\tau-1)\cdot k_\tau\right)
\end{equation}
where $k_\tau$ is the number of $\tau$ cliques the focal vertex belongs to.

\section{Perfectly dissasortative matrices}
\label{appendix:B}

Our definition of perfectly dissasortative symmetric matrices includes a zero diagonal $e_{ii}=0$ whose off diagonal elements $e_{ij}$ sum to a given value $\sum_{ij}e_{ij}=q_i$ such that all entries are positive. Once the structure of a given matrix has been found, the excess joint degree labels can be arranged to fit the required mixing patterns. Extending the example from Eq \ref{eq:3x3}, consider a 5x5 matrix $M_\tau$ with zero diagonal entries and with off-diagonal elements $e_{i,j}$ given by
\begin{subequations}
\begin{align}
    e_{1,2} =&\ {q_1 + q_2 + q_3 + q_4 - q_5} \\
    e_{1,3} =&\  {q_1 + q_2 + q_3 - q_4 - q_5} \\
    e_{1,4} =&\  {q_1 - q_2 - q_3 + q_4 + q_5} \\
    e_{1,5} =&\  {q_1 - q_2 - q_3 - q_4 + q_5} 
\end{align}
\end{subequations}
\begin{subequations}
\begin{align}
    e_{2,1} =&\   {q_1+q_2 + q_3 + q_4 - q_5} \\
    e_{2,3} =&\   { q_1+q_2 + q_3 - q_4 + q_5} \\
    e_{2,4} =&\   {-q_1+q_2 - q_3 + q_4 - q_5}\\
    e_{2,5} =&\   {- q_1+q_2 - q_3 - q_4 + q_5}
\end{align}
\end{subequations}
\begin{subequations}
\begin{align}
    e_{3,1} =&\   {q_1 + q_2+q_3 - q_4 - q_5}\\
    e_{3,2} =&\   { q_1+q_2 + q_3 - q_4 + q_5} \\
    e_{3,4} =&\   {- q_1 - q_2 + q_3 + q_4 - q_5} \\
    e_{3,5} =&\  { -q_1 - q_2 + q_3 + q_4 + q_5} 
\end{align}
\end{subequations}
\begin{subequations}
\begin{align}
    e_{4,1} =&\   {q_1 - q_2 - q_3+q_4 + q_5}\\
    e_{4,2} =&\  {- q_1+q_2 - q_3 + q_4 - q_5}\\
    e_{4,3} =&\   {- q_1 - q_2 + q_3 + q_4 - q_5} \\
    e_{4,5} =&\   { q_1 + q_2 + q_3 + q_4 + q_5} 
\end{align}
\end{subequations}
\begin{subequations}
\begin{align}
    e_{5,1} =&\  {q_1 - q_2 - q_3 - q_4+q_5}\\
    e_{5,2} =& \ {- q_1+q_2 - q_3 - q_4 + q_5}\\
    e_{5,3} =&\   { -q_1 - q_2 + q_3 + q_4 + q_5}\\
    e_{5,4} =& \   { q_1 + q_2 + q_3 + q_4 + q_5}  
\end{align}
\end{subequations}
The matrix $e_{\tau}=0.25M$ is a perfectly dissasortative mixing matrix that satisfies the constraints on the system given by Eqs \ref{eq:constraints}. Following this prescription, only matrices with odd order, $n$, can be found. This is because, for odd $n$ the number of positive elements per row is $n-1$, which must be even for pairs of elements whose net sum is zero to be found. It is unclear if other methods of generating perfectly dissasortative matrices that satisfy Eqs \ref{eq:constraints} exist.

\end{widetext}

\end{document}